\documentclass[
 reprint,
 prl,
]{revtex4-1}

\usepackage{dcolumn}
\usepackage{bm}
\usepackage{xcolor}
\usepackage{graphicx,amsmath,amsfonts}
\usepackage{hyperref}

\newcommand{\ma}{m_{\rm a}}
\newcommand{\rhoa}{{\rho_{\rm a}}}
\newcommand{\gag}{g_{a\gamma}}
\newcommand{\Dlens}{{\Delta\,\theta_{\rm a,lens}}}
\newcommand{\Da}{{\Delta\,\theta_{\rm a}}}

\begin{document}

\preprint{APS/123-QED}

\title{Searching for axion-like particles under strong gravitational lenses}

\author{Aritra Basu$^{1}$}
\email{Email : aritra@physik.uni-bielefeld.de}
\author{Jishnu Goswami$^1$}
\author{Dominik J. Schwarz$^1$}
\author{Yuko Urakawa$^{1,2}$}
\affiliation{%
$^1$Fakult\"at f\"ur Physik, Universit\"at Bielefeld, Postfach 100131, 33501 Bielefeld, Germany\\
$^2$Department of Physics and Astrophysics, Nagoya University, Chikusa, Nagoya 464-8602, Japan
}%

%
%

\date{\today}

\begin{abstract}
We establish strong gravitational lens systems as robust probes of axion-like	
particles (ALPs) --- a candidate for dark matter. A tiny interaction of
photons with ALPs induces birefringence.  Multiple images of
gravitationally lensed polarised objects allow differential
birefringence measurement, alleviating systematics and astrophysical
dependencies. We apply this novel method to the lens system
CLASS\,B1152+199 and constrain ALP-photon coupling $\le 9.2\times
10^{-11}\, {\rm GeV}^{-1} \textrm{ to } 7.7\times 10^{-8}\, {\rm
GeV}^{-1}$ ($95\%$ C.L.) for ALP mass between $3.6\times 10^{-21}
\,{\rm eV}$ and $4.6 \times 10^{-18} \,{\rm eV}$. A larger sample will
improve the constraints.

\end{abstract}

\maketitle

\paragraph*{Introduction.---}
Gravitational effects revealed that a quarter of the energy in the Universe is
contained in \textit{dark matter}.  Its nature remains elusive, as it interacts
only weakly with visible matter.  Various particles beyond the Standard Model
have been proposed as dark matter candidates \cite{berto18}. {\it Axions}
\cite{peccei77,wilcz78} and {\it axion-like particles} (ALPs) are very
promising candidates  \cite{Kim:1979if,shifm80,Dine:1981rt,Zhitnitsky:1980}.
The allowed mass range for ALPs spans tens of orders of magnitude and massive
efforts are underway to search for their signatures through a multitude of
approaches \cite[e.g.,][]{graham15, hloze15, irast18, sigl18}. 

A promising direction of ALP searches focuses on its parity-violating
interaction with photons through the coupling $\gag$, causing the left- and
right-handed circular polarisations of light to propagate at different
velocities in the ALP field --- the {\it birefringence} phenomenon
\cite{harar92, carro98}.  Consequently, the plane of polarisation of linearly
polarised light is rotated with respect to the plane at emission and the amount
of rotation $\Delta\,\theta_{\rm a}$ depends on $\gag$.  Since the ALP field
oscillates in time with period $T_{\rm a} = 2\pi/m_{\rm a}$, where $m_{\rm a}$
denotes the ALP mass, $\Delta\,\theta_{\rm a}$ also oscillates and thus allows
to measure $\ma$.

ALP induced birefringence is achromatic and $\Delta\,\theta_{\rm a}$ can be
measured through observations of suitable linearly polarised astrophysical
systems. For example, indirect linear polarisation at a wavelength of
$1.6\,\mu$m caused by scattered light from the parent star in a proto-planetary
disk \cite{fujit19}, multi-epoch observations at 2\,cm of inherently linearly
polarised synchrotron emission from knots in parsec-scale jets of active
galactic nuclei (AGN) \cite{ivano19}, and $E$-mode polarisation of the cosmic
microwave background radiation \cite{fedde19, sigl18} have been used to
constrain ultra-light ALPs. Besides birefringence, alternative approaches have
also been used, such as, interconversion between photons and ALPs in the
presence of background magnetic fields giving rise to modulations of X-ray
spectrum of AGN in a cluster \cite{berg17,ayad19} and probing gravitational
effects of the oscillating ALP field in the pulse time of arrival from pulsars
in a pulsar timing array \cite{khmeln13,poray18}.

In this work we establish strong gravitational lens systems with polarised
sources as a powerful new probe of ultra-light ALPs via birefringence.
Established astrophysical probes of ALPs heavily rely on assumptions and
modelling, or are limited by instrumental offset and sensitivity
\cite{sigl18,ivano19,fujit19}, and are often sensitive to measure either
$g_{a\gamma}$ \cite{berg17} or $m_{\rm a}$ \cite{poray18}.  Broadly speaking,
they address different aspects --- do ALPs exist, and, whether they are the
dark matter. In this work, we show that spectro-polarimetric observations at
GHz-frequencies of strong gravitational lens systems that produce multiple images
of a polarised source provide a unique advantage for probing dark matter in the
form of ultra-light ALPs. Multiple images from lensing allow for performing
differential polarisation angle measurements which alleviates instrumental
offsets and does not rely on modelling the intrinsic astrophysics of the
system. Differential measurements facilitates firm estimation of $m_{\rm a}$
and $g_{a\gamma}$.

\paragraph*{Differential measurements.---}
Measuring the birefringence angle is faced with two challenges --- (1) accuracy
of instrumental polarisation angle calibration, and, (2) additional chromatic
birefringence introduced by Faraday rotation when linearly polarised signals
traverse through magnetised plasma. Although current instruments are capable of
measuring the angle of polarisation to a fraction of a degree, the accuracy of
absolute angle measurement is limited to a few degrees due to the accuracy to
which the polarisation angle of astronomical calibrators are known. The second
challenge, Faraday rotation, depends on the photon frequency $\nu$ wherein the
plane of polarisation is given as $\theta(\nu) = \theta_0 +
\textrm{RM}\,(c/\nu)^2$.  Here, $\theta_0$ is the angle of the plane of linear
polarisation of the polarised source, and, RM is the Faraday rotation measure,
the integral of the magnetic field component parallel to the line of sight
weighted by free-electron number density. The effects of Faraday rotation are
vastly reduced by performing observations at high frequencies ($> 100$~GHz).
However, at these frequencies very few astrophysical systems give rise to
substantial linearly polarised emission.

In most astrophysical sources, linearly polarised emission at few GHz
frequencies directly originates from the synchrotron mechanism.  However, in
these frequencies, Faraday rotation introduces complicated frequency variation
of the linear polarisation parameters, Stokes $Q$ and $U$, when polarised
signal propagates through turbulent magnetised media \cite{sokol98}.  These
frequency variations are captured by polarisation measurements performed over
large bandwidths, and are robustly modelled by applying the technique of Stokes
$Q,U$ fitting. Stokes $Q,U$ fitting enables determination of the Faraday
rotation-corrected polarisation angle $\theta_0$ emitted by a source as one of
the fitted parameters \cite{sulli12,sulli17}.  Since the birefringence induced
by the interaction of photons with ALP field is achromatic, the measured
$\theta_0 = \theta_{\rm qso} + \Delta\,\theta_{\rm a} + \delta\,\theta_{\rm
cal}$. $\theta_{\rm qso}$ is the intrinsic polarisation angle of a linearly
polarised source, e.g., a quasar, and, $\delta\,\theta_{\rm cal}$ is the offset
angle due to improper calibrations.  For observations of a quasar along a
single line of sight, it is unfeasible to determine $\Delta\,\theta_{\rm a}$
without the knowledge of $\theta_{\rm qso}$ and $\delta\, \theta_{\rm cal}$.

Strong gravitational lensing of polarised objects yields a unique advantage in
mitigating the unknown $\theta_{\rm qso}$ and $\delta\, \theta_{\rm cal}$.
Gravitational lensing offers the opportunity to simultaneously observe
time-separated emission from a source, due to gravitational time delay, as lensed
images.  The time-separated images encode the time variation of the oscillating
ALP field at the emitting source and therefore, the polarisation plane of each
lensed images undergo different amount of birefringence.  Comparing the
polarisation angle of lensed images provide information on $\Delta\,
\theta_{\rm a}$. Polarised signals which were emitted by the source at initial
times $t_i$ with $i= {\rm A}, {\rm B}$ for the two images A and B are observed
simultaneously at time $t_{\rm obs}$. The gravitational time delay observed on
Earth $\Delta t_{\rm obs} = |t_{\rm A} - t_{\rm B}|(1 + z_{\rm qso})$, where
$z_{\rm qso}$ is the redshift of the lensed quasar. By performing Stokes $Q,U$
fitting separately for each image, the polarisation angles of image A and B,
$\theta_{\rm 0, A}$ and $\theta_{\rm 0, B}$, can be obtained. Both $\theta_{\rm
0, A}$ and $\theta_{\rm 0, B}$ contain the same $\theta_{\rm qso}$ and $\delta
\theta_{\rm cal}$, but different rotation angles, $\Delta \theta_{\rm a,A}$ and
$\Delta \theta_{\rm a,B}$, arising from the time separation.  Therefore, the
differential angle $\Delta\, \theta_{\rm a, lens} =  \theta_{\rm 0, A} -
\theta_{\rm 0, B} = \Delta\, \theta_{\rm a,A} - \Delta\,\theta_{\rm a,B}$ does
not depend on $\theta_{\rm qso}$ or $\delta\theta_{\rm cal}$.  The significance
of the result is determined by statistical measurement noise. The birefringence
in a similar setup was considered to explore the anomalous coefficient of
axion strings \cite{agraw19} and for detecting cosmic axion background using
polarised pulsars \cite{liu20}.

\paragraph*{Differential birefringence.---}
We consider the Lagrangian density for an ALP field $a$ given as,
\begin{equation}
    \mathcal{L} = -\frac 14 F_{\mu\nu}F^{\mu\nu} - \frac 12 \partial_\mu a \partial^\mu a + \frac{g_{a\gamma}}{4}a F_{\mu\nu}\widetilde{F}^{\mu\nu} - \frac 12 m_{\rm a}^2 a^2,
\label{eq:lagrange}
\end{equation}
where $F_{\mu\nu}$ denotes the electromagnetic field strength tensor and
$\widetilde{F}^{\mu\nu}$ is its dual.  We use the Heaviside-Lorentz system with
$\hbar = c = 1$. The equation of motion for $a$ is given by the Klein-Gordon
equation, solved as,
\begin{equation}
    a(t, x^i) = \frac{\sqrt{2\, \rho_{\rm a}(x^i)}}{m_{\rm a}}\, \sin\left(m_{\rm a}\, t + \delta(x^i) \right).
\label{eq:eom}
\end{equation}
Here, $x^i$ represents the three spatial coordinates, $\rhoa$ is the energy
density of the ALP field, and $\delta$ is the phase.  Inhomogeneities of the
ALP field are encoded in the spatial dependencies of $\rhoa$ and $\delta$, with
$\delta$ being constant within patches of size of the de Broglie wavelength
$\lambda_{\rm dB}$. 

The parity-violating coupling term $a F\widetilde{F}$ gives rise to
birefringence.  As the ALP field oscillates, the birefringence angle also
oscillates in time.  When temporal and spatial variations of the ALP field are
much smaller than the frequency of the photons propagating in the ALP field,
satisfying $10^{-16}(m/10^{-22}\,{\rm eV}) ({\rm GHz}/\nu) \ll 1$, the rotation
angle is given as \cite{harar92, sigl18, fedde19,schwarz20},
\begin{equation}
    \Delta\,\theta_{\rm a} = \frac 12 \, g_{a\gamma}\, \left[ a(t_{\rm obs}, x_{\rm obs}^i) - a(t_{\rm em}, x_{\rm
 em}^i)\right].
\label{eq:angle}
\end{equation}
The subscripts `obs' and `em' indicate the ALP field at observation and at
photon emission, respectively. Thus, in order to infer $\gag$, the field values
at emission region and at observations has to be known or assumed
\cite{ivano19}.

For $\Da$ measured towards two gravitationally lensed images A and B, the field
$a(t_{\rm obs}, x_{\rm obs}^i)$ is the same. In such a case, the differential
birefringence angle $\Dlens = \Delta\,\theta_{\rm a, A} - \Delta\,\theta_{\rm
a, B}$ depends only on the properties of ALP field in the emitting region and
is not affected by the space-time curvature of the gravitational lens
\cite{schwarz20}.  Thus, using Eqs.~\eqref{eq:eom} and \eqref{eq:angle}, 
\begin{equation}
    \Delta\,\theta_{\rm a,lens} = K \sin\!\left[\frac{m_{\rm a} \Delta t}{2} \right] \sin\left(m_{\rm a} t_
{\rm em} + \delta_{\rm em} - \pi/2 \right).
\label{eq:Dlens}
\end{equation}
Here, $K$ in normalized units is,
\begin{equation}
    K = 10^\circ\!  \left[ \frac{\rho_{\rm a,em} }{20\,{\rm GeV \, cm^{-3}} }\right]^{1/2}\!\!\! 
    \frac{g_{a\gamma}}{10^{-12}\,{\rm GeV}^{-1}} 
    \left[\frac{m_{\rm a}}{10^{-22}\,{\rm eV}} \right]^{-1}\!\! .
\label{eq:norm}
\end{equation}
We have used $\rho_{\rm a, A} = \rho_{\rm a, B} \equiv \rho_{\rm a, em}$ (the
energy density of the ALP field in the emitting region), and, $t_{\rm em} =
(t_{\rm A} + t_{\rm B})/2$ is the mean time of emission. The phase difference
$\delta_{\rm em}$ is the same for the emitting region.  Thus, the amplitude of
the differential birefringence signal depends on the lensing time delay
measured in the frame of the emitting source, $\Delta \, t = |t_{\rm A} -
t_{\rm B}|$.  In Eq.~\eqref{eq:Dlens}, the $\sin\left(m_{\rm a}\, t_{\rm em} +
\delta_{\rm em} - \pi/2 \right)$ term suggests that $\Dlens$ would also
oscillate with the same time period as that of the ALP field determined by
$\ma$, but shifted in phase by $90^\circ$. With sufficiently sensitive
observations, the oscillation of $\Dlens$ can be measured through regular
monitoring of a lens system and $\ma$ can be measured. 

Based on a measurement of $\Dlens$ at a single epoch, the ratio of the
ALP-photon coupling and ALP mass, $g_{a\gamma}/m_a$, can be inferred for a
given $\rho_{\rm a}$ determined through ancillary measurements. This therefore
provides $\gag$ over a range of $\ma$. The time-scales involved in the
emission, observations, and light propagation determine the range of ALP masses
that can be probed.  For a single epoch, $\Delta\, \theta_{\rm a,lens}$ is
given as, $\Delta\, \theta_{\rm a,lens} = K\, \sin\left(m_{\rm a}\,
\Delta\,t/2\right)/\sqrt{2}$, where the factor $1/\sqrt{2}$ originates from the
root mean square of the oscillating ALP field with a random phase.  Note that,
in the case when $T_{\rm a} \gg \Delta t$, the values of the field for image A
and B would be close to each other and the differential birefringence would be
negligible.  Therefore, the lensing time delay determines the minimum ALP mass
($\ma = 2\pi/\Delta t$) to which a single epoch observation is sensitive.  The
maximum ALP mass is determined by the averaging time $t_{\rm avg}$ of
observations in the frame of the lensed object. This is because, field
oscillations on smaller time-scales due to ALP with $m_{\rm a} > 2\,\pi/t_{\rm
avg}$ will be averaged out in the observations.

When a lens system is regularly observed over a period of time $t_{\rm monitor}
\gg \Delta t$, the accessible mass range increases towards lower $\ma =
2\pi/t_{\rm monitor}$. With sufficiently sensitive observations during the
course of the monitoring, the sinusoidal oscillation of $\Dlens$ should be
observed if $\ma$ is smaller than the corresponding time over which data are
averaged in each of the observations. In the event of non-detection, limits
obtained on $\Dlens$ could extend the ruled out parameter space for ALPs as
compared to that obtainable for a single epoch observation.  A long monitoring
program corresponds to the limit $\ma\Delta t \ll 1$, and from
Eq.~\eqref{eq:Dlens}, $\Dlens$ can be constrained using the relation $\Dlens =
K\,\ma\,\Delta t/2\,\sqrt{2}$.

\begin{figure} 
	\includegraphics[width=8cm, trim=10mm 50mm 10mm 10mm, clip]{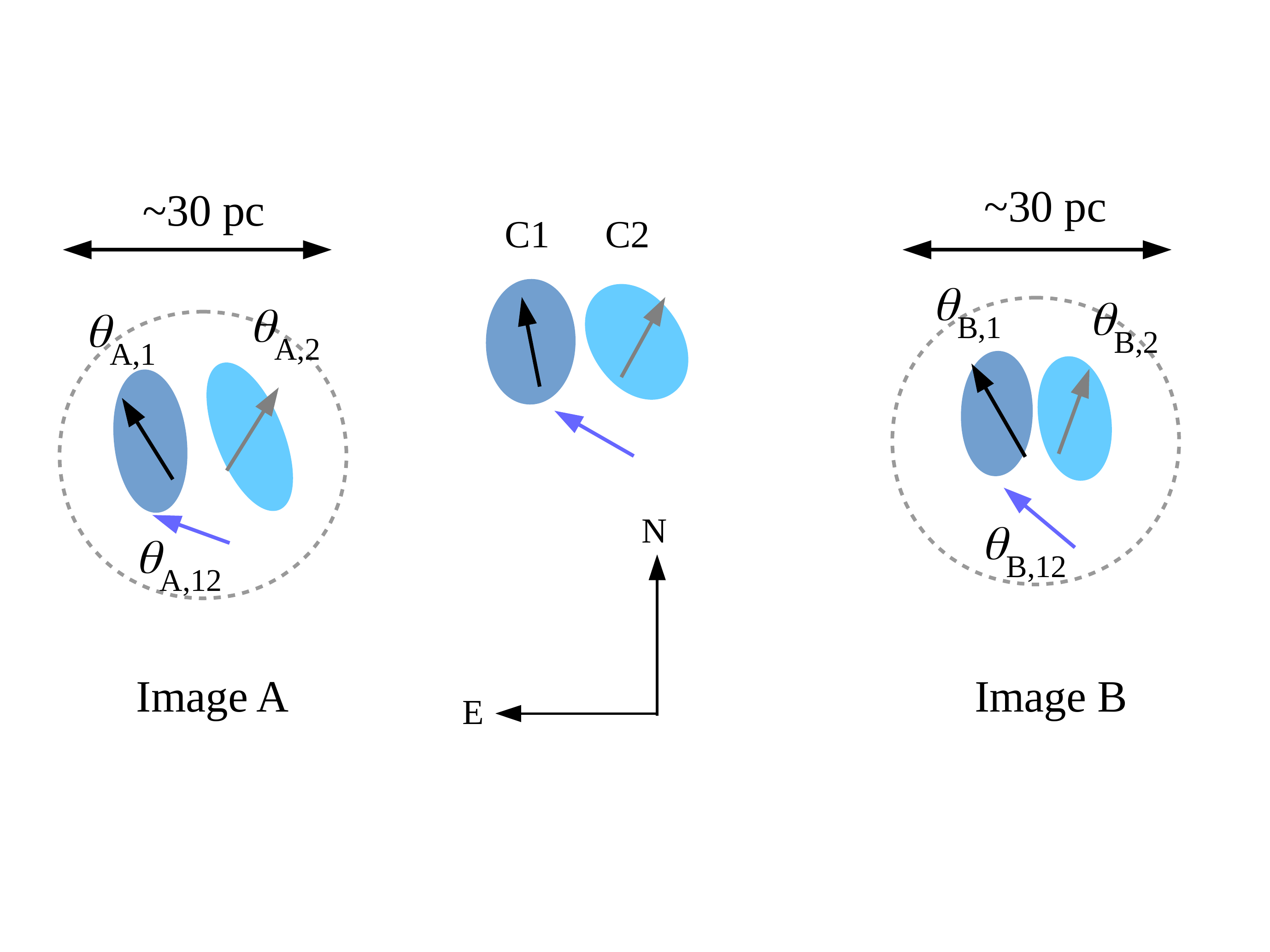}
\caption{Schematic of polarisation angle orientations for the two lensed images
	of the background quasar in CLASS\,B1152+199. The angles $\theta_{{\rm
	A},i}$ and $\theta_{{\rm B},i}$ ($i =$ 1, 2, and, 12) are used for our
	calculations.  The presented angles are measured towards East from
	North of the coordinate axes shown here. The blobs marked as `C1' and
	`C2' depict the two polarised components.  The dashed circles show a
	schematic spatial resolution of $\sim 30$~pc at the distance of the
	quasar.  The relative shapes and orientations of the components in the
	lensed images are different to indicate shearing due to lensing. The
	extent of shearing of the components and their angles are exaggerated
	for representation and are not to scale.}
\label{fig:cartoon}
\end{figure}

\paragraph*{Constraint from CLASS\,B1152+199.---}
The lens system CLASS\,B1152+199 is a highly suitable candidate for probing
ALPs and has been studied in detail in the literature providing relevant
information.  CLASS\,B1152+199 is a galaxy gravitational lens system wherein a
foreground star-forming galaxy at redshift $z_{\rm gal}=0.439$ lenses a
background linearly polarised quasar at redshift $z_{\rm qso}=1.019$
\cite{myers99,rusin02}.  Using an isothermal sphere mass distribution model for
the lensing galaxy, the best-fit time delay $\Delta\,t_{\rm obs}$ is estimated
to be 27.8 days \cite{rusin02}, which corresponds to $\Delta\,t = 13.3$~day in
the frame of the quasar \footnote{Note that, \cite{rusin02} computed $\Delta
t_{\rm obs} =$ 39.7 day for the assumed Hubble-Lema\^itre constant $H_0 =
100~\rm km\,s^{-1}\,Mpc^{-1}$. Here, we have corrected the estimated time delay
using $H_0 = 67.4~\rm km\,s^{-1}\,Mpc^{-1}$ obtained from the {\it Planck}
satellite's data \cite{planck18VI} and used $\Delta t_{\rm obs} =$ 27.8~day in
our calculations.}.

Broad-band polarisation observations of this system, covering the frequency
range 1 to 8~GHz, were performed using the Karl G.  Jansky Very Large Array
(VLA) to constrain the magnetic field geometry in the lensing galaxy
\cite{mao17}. The background quasar is lensed into two images, denoted by image
A and B, separated by 1.56 arcsec. Both images were detected in the VLA
observations, and their polarisation spectra were independently fitted using
the technique of Stokes $Q,U$ fitting \cite{mao17}.  It was found that the
polarised emission of each of the images are composed of two polarisation
emitting components which undergo Faraday rotation and depolarisation in the
magnetised plasma of the lensing galaxy.  These polarised components, denoted
as component 1 and 2, remained unresolved in the VLA observations at up to 0.5
arcsec \cite{mao17}. From milli-arcsec resolution observations \cite{rusin02},
these polarised components are physically separated by $<30$~pc.  The fitted
angle of polarisation of each of the polarised components in the lensed images
are presented in Table~\ref{tab:angle}. We denote the Faraday
rotation-corrected angles with corresponding image and component as subscripts,
i.e., $\theta_{I, C}$ where $I = {\rm A, B}$ represents the lensed images and
$C=1,2$ corresponds to the polarised components.  A pictorial representation of
the different polarisation angles is shown in Fig.~\ref{fig:cartoon}.

\begin{table} 
\centering
\caption{Best-fit intrinsic polarisation angle of the polarised emission 
components in CLASS\,B1152+199 taken from table~1 of \cite{mao17}.}
\centering
  \begin{tabular}{@{}|l|l|l|@{}}
 \hline
Image  &  Component 1 & Component 2 \\
 \hline
A	  & $\theta_{\rm A,1} = 32^\circ\pm1^\circ$ &  $\theta_{\rm A,2}= -38^\circ\pm4^\circ$ \\
\hline
B	  & $\theta_{\rm B,1} = 30^\circ\pm10^\circ$ &  $\theta_{\rm B,2} = -20^\circ\pm20^\circ$ \\
\hline
\end{tabular}
\label{tab:angle}
\end{table}

Presence of multiple polarised emission components in each of the lensed images
of the background quasar provides the advantage of measuring three independent
angle differences for the two lines of sight, i.e, two angle difference of the
individual polarised components $\Delta\,\theta_{\rm 1} = \theta_{\rm A, 1} -
\theta_{\rm B, 1}$ and $\Delta\,\theta_{\rm 2} = \theta_{\rm A, 2} -
\theta_{\rm B, 2}$, and the difference of relative angles between the two
polarised components $\Delta\,\theta_{\rm 12} = (\theta_{\rm A, 1} -
\theta_{\rm A, 2}) - (\theta_{\rm B, 1} - \theta_{\rm B, 2})$. Each of the
angle difference measures $\Delta\,\theta_{\rm a,lens}$, and the combined mean
provides the net angle difference between the two lines of sight. This follows
from the fact that emission from the components within 30~pc is comparable to
$\lambda_{\rm dB}$ for ALP dark matter with $\ma \sim 10^{-20}$~eV and velocity
dispersion $\sim 100~\rm km\,s^{-1}$.

Differential angle measurements do not depend on the choice of the coordinate
system, except for the sense in which the difference is calculated. Here, we
always measure angle differences with respect to image A. Because of
periodicity of angles, we apply directional statistics \cite{fisher93,mardia00}
to compute the mean $\Dlens$. As there are only three independent measures of
birefringence angles, $\Delta\,\theta_1, \Delta\,\theta_2$ and
$\Delta\,\theta_{12}$, we have performed a Monte-Carlo simulation with 50,000
realizations of $\theta_{\rm A,1}, \theta_{\rm A,2}, \theta_{\rm B,1}$ and
$\theta_{\rm B,2}$, each of which is drawn from a modified Von Mises
probability distribution function (PDF) of the form, 
\begin{equation}
    {\rm PDF}(\theta) = \frac{1}{I_0(\kappa)}\, {\rm e}^{\kappa\, \cos\left[2\, (\theta - \mu)\right]},~\textrm{for}\,-\frac{\pi}{2} \le \theta \le +\frac{\pi}{2},
\label{eq:vonMises}
\end{equation}
to account for the fact that measurements of synchrotron polarisation are
sensitive to the orientation of the polarisation plane and not the direction.
$\mu$ is the measured angle, $\kappa$ is the concentration parameter given by
the inverse square of the respective error of the measured angle in
Table~\ref{tab:angle} expressed in radians, and $I_0$ is the modified Bessel
function of order zero. 

\begin{figure} 
\includegraphics[width=8cm]{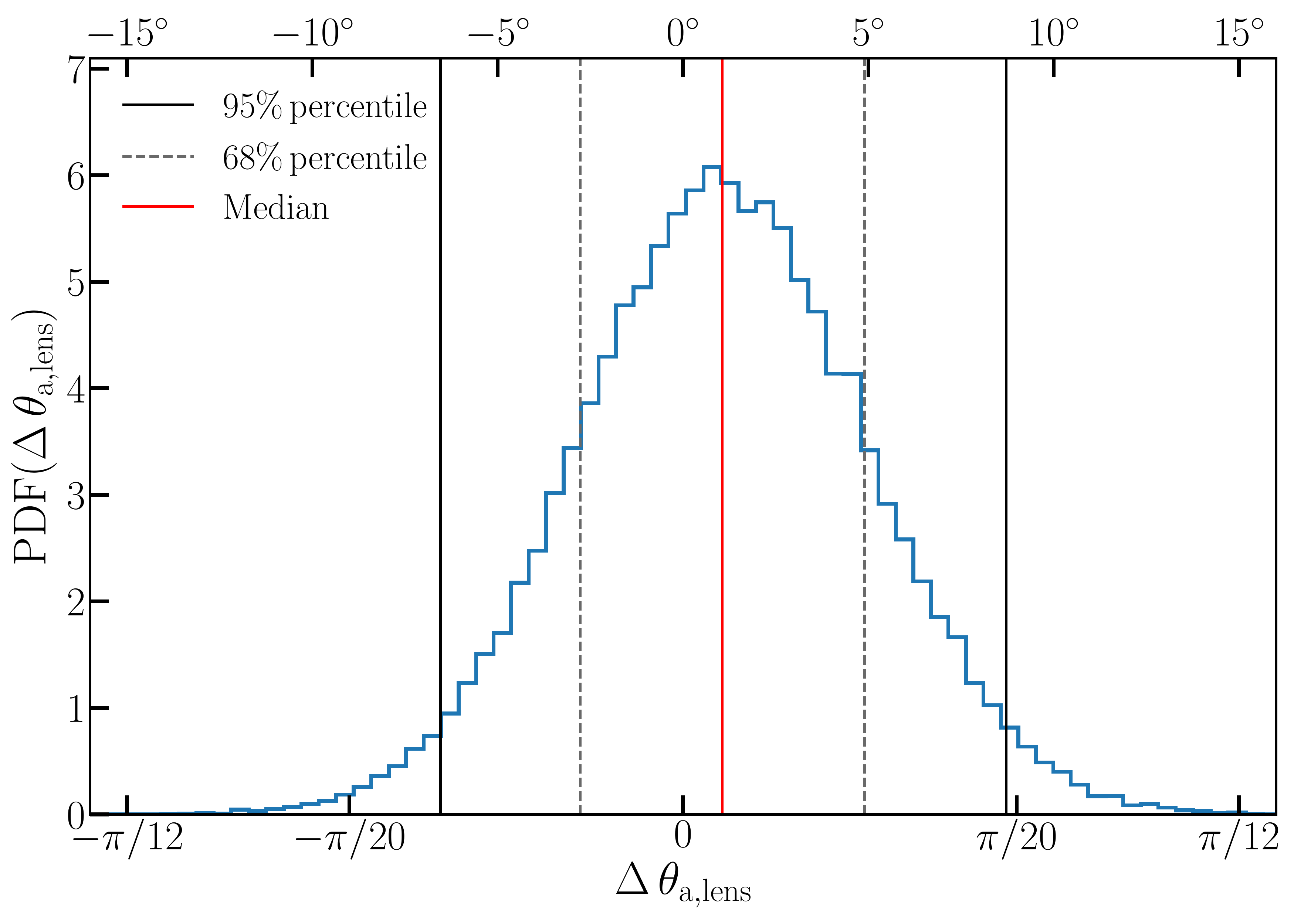}
\caption{Distribution of $\Delta\,\theta_{\rm a,lens}$ calculated from modified
	Von Mises distribution of angles presented in Table~\ref{tab:angle}.
	The red line shows the location of the median of the distribution, and
	the solid and the dashed lines are the 95\% and 68\% percentile
	confidence intervals around the median.} \label{fig:DelthetaPDF}
\end{figure}

Fig.~\ref{fig:DelthetaPDF} shows the PDF of $\Delta \theta_{\rm a,lens} =
\langle \Delta \theta_i\rangle$, where $i = 1, 2$ and 12, computed using
weighted circular mean from random realizations of $\theta_{\rm A,1},
\theta_{\rm A,2}, \theta_{\rm B,1}$ and $\theta_{\rm B,2}$. The inverse of the
dispersion ($\sigma_{\Delta\theta_i}$) of $\Delta\,\theta_1, \Delta\,\theta_2$
and $\Delta\,\theta_{12}$ distributions were used as their corresponding
weights, $w_i = \sigma_{\Delta\theta_i}$. The distribution of $\Delta\,
\theta_{\rm a,lens}$ is well represented by a normal distribution. We used the
median value of the distribution and obtain $\Delta\, \theta_{\rm a,lens} =
1.04_{-1.80}^{+3.90}$ deg (68\% confidence interval) and $\Delta\,
\theta_{\rm a,lens} = 1.04_{-5.59}^{+7.67}$ deg (95\% confidence interval).
These provide the upper limit $|\Delta\, \theta_{\rm a,lens}| < 4.94^\circ$ at
68\% confidence level (C.L.) and $< 8.71^\circ$ at 95\% C.L.  If uniform
weighting ($w_i = 1$) is used instead, the bounds on $|\Dlens|$ improves by
about $1^\circ$.

A part of $\Delta\,\theta_{\rm a,lens}$ could arise due to slight differences
of the geodesics along which the polarised light travels for the two lensed
images and differential gravitational shearing encountered by the two lensed
images. Both these effects are of the order of the lensing deflection angle
\cite{lyuti17}. For CLASS\,B1152+199, the deflection angle is 1.56~arcsec,
significantly lower than the measured error for the angle difference between
the two images. Hence, the contributions of different light travel paths and
shearing in $\Delta\,\theta_{\rm a,lens}$ are negligible.  Further, the total
radio continuum intensity of CLASS\,B1152+199 at 8.46\,GHz was monitored using
the VLA spanning over 7~months with an average observation spacing of 3.5~days
\cite{rumba15}. No time variability in the intensities of the two lensed images
were observed. This strongly suggests that any changes in the polarisation
angles between the polarised components of the two lensed images is unlikely to
be due to intrinsic time variability of the lensed quasar.

Using the estimated limit on $\Delta\,\theta_{\rm a,lens}$, $\gag$ can be
constrained using Eq.~\eqref{eq:Dlens}, where $\rho_{\rm a, em}$ is the only
unknown quantity. It can be assumed within informed range when ALPs are the
dark matter (see below). For the CLASS\,B1152+199 system,  $\rho_{\rm a ,em} =
20~{\rm GeV\,cm^{-3}}$ is assumed for the host galaxy of the quasar.  A factor
of 10 offset in the assumed $\rho_{\rm a,em}$ would lead to systematic offset
by a factor of $\sim3$ in the values of $\gag$.

\begin{figure} 
\includegraphics[width=8cm]{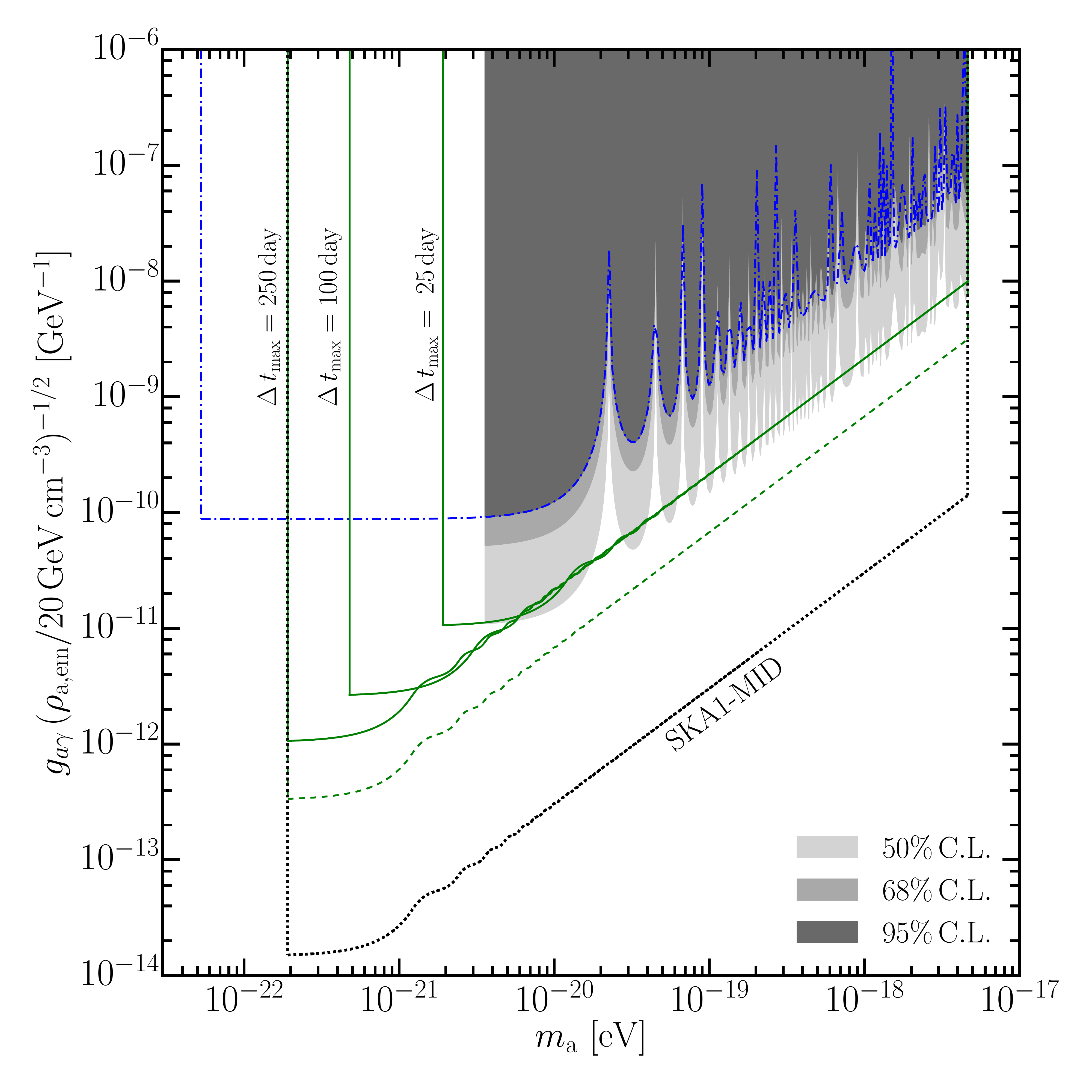}
\caption{Bounds on $g_{a\gamma}$ obtained from 
	CLASS\,B1152+199 is shown as the shaded gray areas. The dash-dotted
	blue line shows the corresponding constraint that can be achieved if
	observations of CLASS\,B1152+199 are sampled over $\sim 5$ year. The
	solid and dashed green lines show the constraint achievable for a
	sample of 100 and 1000 gravitational lens systems, respectively (see
	text).  The various $\Delta\,t_{\rm max}$ values for maximum time delay
	in the frame of the lensed source represent the respective constraints.
	The dashed green line is for $\Delta\,t_{\rm max} = 250$~day. The
	dotted black line shows the parameter space that can be probed by the
	SKA1-MID.}
\label{fig:constraint}
\end{figure}

In Fig.~\ref{fig:constraint}, we present our constraint $g_{a\gamma} \leq 9.2
\times 10^{-11}$ $\left(20\,{\rm GeV\, cm}^{-3}/\rho_{\rm a,
em}\right)^{1/2}~{\rm GeV}^{-1}$ to $7.7 \times 10^{-8}$ $\left(20\,{\rm GeV\,
cm}^{-3}/ \rho_{\rm a, em}\right)^{1/2}~{\rm GeV}^{-1}$ at 95\% C.L. for ALP in
the mass range $3.6 \times 10^{-21}~{\rm eV} \le m_{\rm a} \le 4.6\times
10^{-18}~{\rm eV}$.  As discussed earlier, the maximal ALP mass these
observations are sensitive to is determined by the observations averaging time
of 30~minute \cite{mao17}, i.e., $t_{\rm avg} = 14.86$~minute corresponding to
$m_{\rm a} = 4.6 \times 10^{-18}$~eV.  The minimal $m_{\rm a}$ probed is
determined by the time difference $\Delta t$ over which the emission from the
lensed images originate.  For CLASS\,B1152+199, $\Delta t = 13.3$~day
corresponds to $m_{\rm a} = 3.6 \times 10^{-21}$~eV. 

For a measurement at a single epoch, $m_{\rm a}$ remains unknown, even if a
detection of $\Delta \, \theta_{\rm a, lens}$ would be obtained through
sensitive follow-up observations of CLASS\,B1152+199.  Until such data are
available, the blue dashed-dotted line in Fig.~\ref{fig:constraint} shows the
parameter space that can be probed by the CLASS\,B1152+199 system at 95\% C.L.
if observations with similar sensitivity as those of the observations used here
are sampled over about 5 year (see above).

\paragraph*{Statistical sample.---}
Constraints on $\gag$ probed using gravitational lensing can be vastly improved
for a statistical sample. As this method makes minimal assumptions on
astrophysical processes, the sample can consist of any type of lensed system
where the lensed object is polarised.  As the signal is measured via simple
differences, $\Dlens$ computed for individual lens systems can be simply
combined to compute the mean differential birefringence angle $\langle |
\Dlens|\rangle$ of a sample of $N$ lens systems and reducing the uncertainty by
$\mathcal{O}(1/\sqrt N)$.

In a sample of lensing systems, the time delay $\Delta t$ is a random variable
and $\langle | \Dlens|\rangle$ is proportional to $\langle | \sin\left(\ma
\Delta t/2\right)|\rangle$, because the variables $\rho_{\rm a, em}$ and
$\Delta t$ are statistically independent. For a random argument of
$\sin\left(m_{\rm a} t_{\rm em} + \delta_{\rm em} - \pi/2 \right)$, the sample
average $\langle |\sin\left(m_{\rm a} t_{\rm em} + \delta_{\rm em} - \pi/2
\right)|\rangle = 2/\pi$. Assuming a uniform PDF for $\Delta t \in [0, \Delta
t_{\rm max}]$, $\langle | \Dlens|\rangle$ is given by,
\begin{eqnarray}
        \langle | \Delta \theta_{\rm a, lens}|\rangle   
        & = &
        \frac{2 g_{a\gamma} \sqrt{2 \langle\rho_{\rm a, em}\rangle}}{\pi m_{\rm a}\Delta t_{\rm max}} \int\limits_0^{\Delta t_{\rm max}} \left| \sin \left(\frac{m_{\rm a} \Delta t}{2} \right)\right| {\rm d}(\Delta t), \nonumber \\ 
 &=& \frac{4 K}{\pi m_{\rm a}\Delta t_{\rm max}}\left(2n + 1 - \cos \zeta \right).
\label{eq:DlensSample}
\end{eqnarray}
Here, $\Delta t_{\rm max}$ is the maximum time delay at the frame of lensed
quasars in the sample of gravitational lens systems, $\langle\rho_{\rm a,
em}\rangle$ is the sample mean energy density of the ALP field \footnote{Here
we have made the approximation $\langle \rho_{\rm a, em}^{1/2} \rangle \approx
\langle \rho_{\rm a, em} \rangle^{1/2}$. This is valid when $\langle \rho_{\rm
a, em}\rangle$ is larger than the dispersion of $\rho_{\rm a, em}$}, and $K$ is
the same as Eq.~\eqref{eq:norm} except that $\rho_{\rm a, em}$ is replaced with
$\langle\rho_{\rm a, em}\rangle$. The angle $\ma \Delta t_{\rm max}/2$ is
expressed as $\ma \Delta t_{\rm max}/2 = n\pi + \zeta$, where $\zeta < \pi$ and
$n \in \mathbb{Z}^+$ because $|\sin x|$ has $n\pi$ periodicity. 

In a statistical sample, the value of the mean birefringence angle depends on
$\langle \rho_{\rm a, em} \rangle$ which can be inferred from dark matter
density measured in elliptical galaxies, which host quasars. The total matter
density within the Einstein's radii for a sample of elliptical galaxies have
been measured through lensing and stellar kinematics \cite{koopm06,lysko18}.
Considering 50\% of the mass in ellipticals is contributed by dark matter
\cite{lysko14,lovel18} and ALPs with $m_{\rm a} > 10^{-24}$~eV comprises the
entire dark matter \cite{hloze15}, we estimate $\langle \rho_{\rm a, em}\rangle
= 25$~GeV\,cm$^{-3}$ with dispersion 14.5~GeV\,cm$^{-3}$.  Thus, the
constraints on $\gag$ that can be obtained from a sample of gravitational lens
systems suffer less systematics as compared to assuming a value of $\rho_{\rm
a, em}$ for a single lens system, which, unless measured for that system, can
differ between individual objects.

The solid (dashed) green lines in Fig.~\ref{fig:constraint} show the prediction
for a sample of 100 (1000) gravitational lens systems having different values
of $\Delta t_{\rm max}$ in the frame of the lensed object for $\langle
\rho_{\rm a, em} \rangle = 20$~GeV\,cm$^{-3}$.  Due to the large sample size,
the sensitivity in measuring the birefringence angle improves by a factor of 10
(32) over the value obtained for CLASS\,B1152+199.  The Phase 1 of the Square
Kilometre Array-MID (SKA1-MID) is expected to detect about $10^5$ strong
gravitational lens systems \cite{mckea15}. Assuming that 5\% of the systems
will be polarised, we compute the parameter space that can be probed to
constrain $g_{a\gamma}$ and is shown by the space above the dotted black line
in Fig.~\ref{fig:constraint}. For all the expected constraints using different
statistical samples, we have extrapolated using the 95\% C.L. bound obtained
for CLASS\,B1152+199 and for the SKA1-MID we used a factor of 10 better
sensitivity as compared to that of the VLA \cite{braun}.

\paragraph*{Future prospects.---}
For the ALP mass range probed in this work, the strongest constrain on
$g_{a\gamma}$ is obtained using an alternate approach of investigating
modulations of the photon spectrum in the X-ray waveband of AGN in a cluster
galaxy due to ALP-photon conversion induced by the cluster's magnetic fields
\cite{berg17}. The upper limit $g_{a\gamma} < 1.4 \times 10^{-12}~{\rm
GeV}^{-1}$ (95\% C.L.) depends on astrophysical parameters, such as, structural
properties of magnetic fields, free-electron density, and a model of the AGN's
X-ray spectrum. The parameter space that can be probed by lensing using the
SKA1-MID will improve upon existing bounds on $g_{a\gamma}$ by almost two
orders of magnitude for the mass range $10^{-22}$ to $10^{-20}$~eV and by up to
an order of magnitude in the mass range $10^{-20}$ to $10^{-19}$~eV
\cite{berg17,sigl18}. 

We have established differential birefringence from strong gravitational
lensing as a new robust probe of ultra-light ALPs.  In contrast to the existing
limit on $g_{a\gamma}$, constraints (and perhaps detection) of ALP from
differential birefringence measurements would be independent of astrophysical
parameters.  Future broad-bandwidth radio polarisation observations of
gravitational lensing systems using sensitive radio frequency observations,
such as using the SKA or the VLA, will play consequential role in our quest to
understand the nature of dark matter.

\begin{acknowledgments}
We thank Pranjal Trivedi, Dietrich B\"odeker, Giuseppe Gagliardi, Jordi Miralda
	Escud{\'e} and Anirban Lahiri for insightful discussions on axion
	physics and ALP birefringence. We acknowledge financial support by the
	German Federal Ministry of Education and Research (BMBF) under grant
	05A17PB1 (Verbundprojekt D-MeerKAT) and from the Deutsche
	Forschungsgemeinschaft (DFG, German Research Foundation) through the
	CRC-TR 211 `Strong-interaction matter under extreme conditions'–
	project number 315477589-TRR 211.  Y.~U. acknowledges support by JSPS
	KAKENHI Grant Numbers JP18H04349 and JP19H01894 and in part by
	YITP-T-19-02.
\end{acknowledgments}


\begin{thebibliography}{41}%
\makeatletter
\providecommand \@ifxundefined [1]{%
 \@ifx{#1\undefined}
}%
\providecommand \@ifnum [1]{%
 \ifnum #1\expandafter \@firstoftwo
 \else \expandafter \@secondoftwo
 \fi
}%
\providecommand \@ifx [1]{%
 \ifx #1\expandafter \@firstoftwo
 \else \expandafter \@secondoftwo
 \fi
}%
\providecommand \natexlab [1]{#1}%
\providecommand \enquote  [1]{``#1''}%
\providecommand \bibnamefont  [1]{#1}%
\providecommand \bibfnamefont [1]{#1}%
\providecommand \citenamefont [1]{#1}%
\providecommand \href@noop [0]{\@secondoftwo}%
\providecommand \href [0]{\begingroup \@sanitize@url \@href}%
\providecommand \@href[1]{\@@startlink{#1}\@@href}%
\providecommand \@@href[1]{\endgroup#1\@@endlink}%
\providecommand \@sanitize@url [0]{\catcode `\\12\catcode `\$12\catcode
  `\&12\catcode `\#12\catcode `\^12\catcode `\_12\catcode `\%12\relax}%
\providecommand \@@startlink[1]{}%
\providecommand \@@endlink[0]{}%
\providecommand \url  [0]{\begingroup\@sanitize@url \@url }%
\providecommand \@url [1]{\endgroup\@href {#1}{\urlprefix }}%
\providecommand \urlprefix  [0]{URL }%
\providecommand \Eprint [0]{\href }%
\providecommand \doibase [0]{http://dx.doi.org/}%
\providecommand \selectlanguage [0]{\@gobble}%
\providecommand \bibinfo  [0]{\@secondoftwo}%
\providecommand \bibfield  [0]{\@secondoftwo}%
\providecommand \translation [1]{[#1]}%
\providecommand \BibitemOpen [0]{}%
\providecommand \bibitemStop [0]{}%
\providecommand \bibitemNoStop [0]{.\EOS\space}%
\providecommand \EOS [0]{\spacefactor3000\relax}%
\providecommand \BibitemShut  [1]{\csname bibitem#1\endcsname}%
\let\auto@bib@innerbib\@empty
\bibitem [{\citenamefont {{Bertone}}\ and\ \citenamefont
  {{Tait}}(2018)}]{berto18}%
  \BibitemOpen
  \bibfield  {author} {\bibinfo {author} {\bibfnamefont {G.}~\bibnamefont
  {{Bertone}}}\ and\ \bibinfo {author} {\bibfnamefont {T.~M.~P.}\ \bibnamefont
  {{Tait}}},\ }\href {\doibase 10.1038/s41586-018-0542-z} {\bibfield  {journal}
  {\bibinfo  {journal} {Nature}\ }\textbf {\bibinfo {volume} {562}},\ \bibinfo
  {pages} {51} (\bibinfo {year} {2018})},\ \Eprint
  {http://arxiv.org/abs/1810.01668} {arXiv:1810.01668 [astro-ph.CO]}
  \BibitemShut {NoStop}%
\bibitem [{\citenamefont {{Peccei}}\ and\ \citenamefont
  {{Quinn}}(1977)}]{peccei77}%
  \BibitemOpen
  \bibfield  {author} {\bibinfo {author} {\bibfnamefont {R.~D.}\ \bibnamefont
  {{Peccei}}}\ and\ \bibinfo {author} {\bibfnamefont {H.~R.}\ \bibnamefont
  {{Quinn}}},\ }\href {\doibase 10.1103/PhysRevLett.38.1440} {\bibfield
  {journal} {\bibinfo  {journal} {Phys. Rev. Lett.}\ }\textbf {\bibinfo
  {volume} {38}},\ \bibinfo {pages} {1440} (\bibinfo {year}
  {1977})}\BibitemShut {NoStop}%
\bibitem [{\citenamefont {{Wilczek}}(1978)}]{wilcz78}%
  \BibitemOpen
  \bibfield  {author} {\bibinfo {author} {\bibfnamefont {F.}~\bibnamefont
  {{Wilczek}}},\ }\href {\doibase 10.1103/PhysRevLett.40.279} {\bibfield
  {journal} {\bibinfo  {journal} {Phys. Rev. Lett.}\ }\textbf {\bibinfo
  {volume} {40}},\ \bibinfo {pages} {279} (\bibinfo {year} {1978})}\BibitemShut
  {NoStop}%
\bibitem [{\citenamefont {Kim}(1979)}]{Kim:1979if}%
  \BibitemOpen
  \bibfield  {author} {\bibinfo {author} {\bibfnamefont {J.~E.}\ \bibnamefont
  {Kim}},\ }\href {\doibase 10.1103/PhysRevLett.43.103} {\bibfield  {journal}
  {\bibinfo  {journal} {Phys. Rev. Lett.}\ }\textbf {\bibinfo {volume} {43}},\
  \bibinfo {pages} {103} (\bibinfo {year} {1979})}\BibitemShut {NoStop}%
\bibitem [{\citenamefont {Shifman}\ \emph {et~al.}(1980)\citenamefont
  {Shifman}, \citenamefont {Vainshtein},\ and\ \citenamefont
  {Zakharov}}]{shifm80}%
  \BibitemOpen
  \bibfield  {author} {\bibinfo {author} {\bibfnamefont {M.}~\bibnamefont
  {Shifman}}, \bibinfo {author} {\bibfnamefont {A.}~\bibnamefont {Vainshtein}},
  \ and\ \bibinfo {author} {\bibfnamefont {V.}~\bibnamefont {Zakharov}},\
  }\href {\doibase https://doi.org/10.1016/0550-3213(80)90209-6} {\bibfield
  {journal} {\bibinfo  {journal} {Nucl. Phys. B}\ }\textbf {\bibinfo {volume}
  {166}},\ \bibinfo {pages} {493 } (\bibinfo {year} {1980})}\BibitemShut
  {NoStop}%
\bibitem [{\citenamefont {Dine}\ \emph {et~al.}(1981)\citenamefont {Dine},
  \citenamefont {Fischler},\ and\ \citenamefont {Srednicki}}]{Dine:1981rt}%
  \BibitemOpen
  \bibfield  {author} {\bibinfo {author} {\bibfnamefont {M.}~\bibnamefont
  {Dine}}, \bibinfo {author} {\bibfnamefont {W.}~\bibnamefont {Fischler}}, \
  and\ \bibinfo {author} {\bibfnamefont {M.}~\bibnamefont {Srednicki}},\ }\href
  {\doibase 10.1016/0370-2693(81)90590-6} {\bibfield  {journal} {\bibinfo
  {journal} {Phys. Lett.}\ }\textbf {\bibinfo {volume} {104B}},\ \bibinfo
  {pages} {199} (\bibinfo {year} {1981})}\BibitemShut {NoStop}%
\bibitem [{\citenamefont {Zhitnitsky}(1980)}]{Zhitnitsky:1980}%
  \BibitemOpen
  \bibfield  {author} {\bibinfo {author} {\bibfnamefont {A.}~\bibnamefont
  {Zhitnitsky}},\ }\href@noop {} {\bibfield  {journal} {\bibinfo  {journal}
  {Sov. J. Nucl. Phys.}\ }\textbf {\bibinfo {volume} {31}},\ \bibinfo {pages}
  {260} (\bibinfo {year} {1980})}\BibitemShut {NoStop}%
\bibitem [{\citenamefont {Graham}\ \emph {et~al.}(2015)\citenamefont {Graham},
  \citenamefont {Irastorza}, \citenamefont {Lamoreaux}, \citenamefont
  {Lindner},\ and\ \citenamefont {van Bibber}}]{graham15}%
  \BibitemOpen
  \bibfield  {author} {\bibinfo {author} {\bibfnamefont {P.~W.}\ \bibnamefont
  {Graham}}, \bibinfo {author} {\bibfnamefont {I.~G.}\ \bibnamefont
  {Irastorza}}, \bibinfo {author} {\bibfnamefont {S.~K.}\ \bibnamefont
  {Lamoreaux}}, \bibinfo {author} {\bibfnamefont {A.}~\bibnamefont {Lindner}},
  \ and\ \bibinfo {author} {\bibfnamefont {K.~A.}\ \bibnamefont {van Bibber}},\
  }\href {\doibase 10.1146/annurev-nucl-102014-022120} {\bibfield  {journal}
  {\bibinfo  {journal} {Annual Review of Nuclear and Particle Science}\
  }\textbf {\bibinfo {volume} {65}},\ \bibinfo {pages} {485–514} (\bibinfo
  {year} {2015})},\ \Eprint
  {http://arxiv.org/abs/https://doi.org/10.1146/annurev-nucl-102014-022120}
  {https://doi.org/10.1146/annurev-nucl-102014-022120} \BibitemShut {NoStop}%
\bibitem [{\citenamefont {{Hlozek}}\ \emph {et~al.}(2015)\citenamefont
  {{Hlozek}}, \citenamefont {{Grin}}, \citenamefont {{Marsh}},\ and\
  \citenamefont {{Ferreira}}}]{hloze15}%
  \BibitemOpen
  \bibfield  {author} {\bibinfo {author} {\bibfnamefont {R.}~\bibnamefont
  {{Hlozek}}}, \bibinfo {author} {\bibfnamefont {D.}~\bibnamefont {{Grin}}},
  \bibinfo {author} {\bibfnamefont {D.~J.~E.}\ \bibnamefont {{Marsh}}}, \ and\
  \bibinfo {author} {\bibfnamefont {P.~G.}\ \bibnamefont {{Ferreira}}},\ }\href
  {\doibase 10.1103/PhysRevD.91.103512} {\bibfield  {journal} {\bibinfo
  {journal} {Phys. Rev. D}\ }\textbf {\bibinfo {volume} {91}},\ \bibinfo {eid}
  {103512} (\bibinfo {year} {2015})},\ \Eprint {http://arxiv.org/abs/1410.2896}
  {arXiv:1410.2896 [astro-ph.CO]} \BibitemShut {NoStop}%
\bibitem [{\citenamefont {Irastorza}\ and\ \citenamefont
  {Redondo}(2018)}]{irast18}%
  \BibitemOpen
  \bibfield  {author} {\bibinfo {author} {\bibfnamefont {I.~G.}\ \bibnamefont
  {Irastorza}}\ and\ \bibinfo {author} {\bibfnamefont {J.}~\bibnamefont
  {Redondo}},\ }\href {\doibase https://doi.org/10.1016/j.ppnp.2018.05.003}
  {\bibfield  {journal} {\bibinfo  {journal} {Progress in Particle and Nuclear
  Physics}\ }\textbf {\bibinfo {volume} {102}},\ \bibinfo {pages} {89 }
  (\bibinfo {year} {2018})}\BibitemShut {NoStop}%
\bibitem [{\citenamefont {{Sigl}}\ and\ \citenamefont
  {{Trivedi}}(2018)}]{sigl18}%
  \BibitemOpen
  \bibfield  {author} {\bibinfo {author} {\bibfnamefont {G.}~\bibnamefont
  {{Sigl}}}\ and\ \bibinfo {author} {\bibfnamefont {P.}~\bibnamefont
  {{Trivedi}}},\ }\href@noop {} {\bibfield  {journal} {\bibinfo  {journal}
  {arXiv e-prints}\ ,\ \bibinfo {eid} {arXiv:1811.07873}} (\bibinfo {year}
  {2018})},\ \Eprint {http://arxiv.org/abs/1811.07873} {arXiv:1811.07873
  [astro-ph.CO]} \BibitemShut {NoStop}%
\bibitem [{\citenamefont {Harari}\ and\ \citenamefont
  {Sikivie}(1992)}]{harar92}%
  \BibitemOpen
  \bibfield  {author} {\bibinfo {author} {\bibfnamefont {D.}~\bibnamefont
  {Harari}}\ and\ \bibinfo {author} {\bibfnamefont {P.}~\bibnamefont
  {Sikivie}},\ }\href {\doibase 10.1016/0370-2693(92)91363-E} {\bibfield
  {journal} {\bibinfo  {journal} {Phys. Lett.}\ }\textbf {\bibinfo {volume}
  {B289}},\ \bibinfo {pages} {67} (\bibinfo {year} {1992})}\BibitemShut
  {NoStop}%
\bibitem [{\citenamefont {Carroll}(1998)}]{carro98}%
  \BibitemOpen
  \bibfield  {author} {\bibinfo {author} {\bibfnamefont {S.~M.}\ \bibnamefont
  {Carroll}},\ }\href {\doibase 10.1103/PhysRevLett.81.3067} {\bibfield
  {journal} {\bibinfo  {journal} {Phys. Rev. Lett.}\ }\textbf {\bibinfo
  {volume} {81}},\ \bibinfo {pages} {3067} (\bibinfo {year} {1998})},\ \Eprint
  {http://arxiv.org/abs/astro-ph/9806099} {arXiv:astro-ph/9806099} \BibitemShut
  {NoStop}%
\bibitem [{\citenamefont {Fujita}\ \emph {et~al.}(2019)\citenamefont {Fujita},
  \citenamefont {Tazaki},\ and\ \citenamefont {Toma}}]{fujit19}%
  \BibitemOpen
  \bibfield  {author} {\bibinfo {author} {\bibfnamefont {T.}~\bibnamefont
  {Fujita}}, \bibinfo {author} {\bibfnamefont {R.}~\bibnamefont {Tazaki}}, \
  and\ \bibinfo {author} {\bibfnamefont {K.}~\bibnamefont {Toma}},\ }\href
  {\doibase 10.1103/PhysRevLett.122.191101} {\bibfield  {journal} {\bibinfo
  {journal} {Phys. Rev. Lett.}\ }\textbf {\bibinfo {volume} {122}},\ \bibinfo
  {pages} {191101} (\bibinfo {year} {2019})},\ \Eprint
  {http://arxiv.org/abs/1811.03525} {arXiv:1811.03525 [astro-ph.CO]}
  \BibitemShut {NoStop}%
\bibitem [{\citenamefont {{Ivanov}}\ \emph {et~al.}(2019)\citenamefont
  {{Ivanov}}, \citenamefont {{Kovalev}}, \citenamefont {{Lister}},
  \citenamefont {{Panin}}, \citenamefont {{Pushkarev}}, \citenamefont
  {{Savolainen}},\ and\ \citenamefont {{Troitsky}}}]{ivano19}%
  \BibitemOpen
  \bibfield  {author} {\bibinfo {author} {\bibfnamefont {M.~M.}\ \bibnamefont
  {{Ivanov}}}, \bibinfo {author} {\bibfnamefont {Y.~Y.}\ \bibnamefont
  {{Kovalev}}}, \bibinfo {author} {\bibfnamefont {M.~L.}\ \bibnamefont
  {{Lister}}}, \bibinfo {author} {\bibfnamefont {A.~G.}\ \bibnamefont
  {{Panin}}}, \bibinfo {author} {\bibfnamefont {A.~B.}\ \bibnamefont
  {{Pushkarev}}}, \bibinfo {author} {\bibfnamefont {T.}~\bibnamefont
  {{Savolainen}}}, \ and\ \bibinfo {author} {\bibfnamefont {S.~V.}\
  \bibnamefont {{Troitsky}}},\ }\href {\doibase 10.1088/1475-7516/2019/02/059}
  {\bibfield  {journal} {\bibinfo  {journal} {JCAP}\ }\textbf {\bibinfo
  {volume} {2019}},\ \bibinfo {eid} {059} (\bibinfo {year} {2019})},\ \Eprint
  {http://arxiv.org/abs/1811.10997} {arXiv:1811.10997 [astro-ph.CO]}
  \BibitemShut {NoStop}%
\bibitem [{\citenamefont {Fedderke}\ \emph {et~al.}(2019)\citenamefont
  {Fedderke}, \citenamefont {Graham},\ and\ \citenamefont
  {Rajendran}}]{fedde19}%
  \BibitemOpen
  \bibfield  {author} {\bibinfo {author} {\bibfnamefont {M.~A.}\ \bibnamefont
  {Fedderke}}, \bibinfo {author} {\bibfnamefont {P.~W.}\ \bibnamefont
  {Graham}}, \ and\ \bibinfo {author} {\bibfnamefont {S.}~\bibnamefont
  {Rajendran}},\ }\href {\doibase 10.1103/PhysRevD.100.015040} {\bibfield
  {journal} {\bibinfo  {journal} {Phys. Rev.}\ }\textbf {\bibinfo {volume}
  {D100}},\ \bibinfo {pages} {015040} (\bibinfo {year} {2019})},\ \Eprint
  {http://arxiv.org/abs/1903.02666} {arXiv:1903.02666 [astro-ph.CO]}
  \BibitemShut {NoStop}%
\bibitem [{\citenamefont {{Berg}}\ \emph {et~al.}(2017)\citenamefont {{Berg}},
  \citenamefont {{Conlon}}, \citenamefont {{Day}}, \citenamefont {{Jennings}},
  \citenamefont {{Krippendorf}}, \citenamefont {{Powell}},\ and\ \citenamefont
  {{Rummel}}}]{berg17}%
  \BibitemOpen
  \bibfield  {author} {\bibinfo {author} {\bibfnamefont {M.}~\bibnamefont
  {{Berg}}}, \bibinfo {author} {\bibfnamefont {J.~P.}\ \bibnamefont
  {{Conlon}}}, \bibinfo {author} {\bibfnamefont {F.}~\bibnamefont {{Day}}},
  \bibinfo {author} {\bibfnamefont {N.}~\bibnamefont {{Jennings}}}, \bibinfo
  {author} {\bibfnamefont {S.}~\bibnamefont {{Krippendorf}}}, \bibinfo {author}
  {\bibfnamefont {A.~J.}\ \bibnamefont {{Powell}}}, \ and\ \bibinfo {author}
  {\bibfnamefont {M.}~\bibnamefont {{Rummel}}},\ }\href {\doibase
  10.3847/1538-4357/aa8b16} {\bibfield  {journal} {\bibinfo  {journal}
  {Astrophys. J.}\ }\textbf {\bibinfo {volume} {847}},\ \bibinfo {eid} {101}
  (\bibinfo {year} {2017})},\ \Eprint {http://arxiv.org/abs/1605.01043}
  {arXiv:1605.01043 [astro-ph.HE]} \BibitemShut {NoStop}%
\bibitem [{\citenamefont {{Ayad}}\ and\ \citenamefont {{Beck}}(2019)}]{ayad19}%
  \BibitemOpen
  \bibfield  {author} {\bibinfo {author} {\bibfnamefont {A.}~\bibnamefont
  {{Ayad}}}\ and\ \bibinfo {author} {\bibfnamefont {G.}~\bibnamefont
  {{Beck}}},\ }\href@noop {} {\bibfield  {journal} {\bibinfo  {journal} {arXiv
  e-prints}\ ,\ \bibinfo {eid} {arXiv:1911.10078}} (\bibinfo {year} {2019})},\
  \Eprint {http://arxiv.org/abs/1911.10078} {arXiv:1911.10078 [astro-ph.HE]}
  \BibitemShut {NoStop}%
\bibitem [{\citenamefont {Khmelnitsky}\ and\ \citenamefont
  {Rubakov}(2014)}]{khmeln13}%
  \BibitemOpen
  \bibfield  {author} {\bibinfo {author} {\bibfnamefont {A.}~\bibnamefont
  {Khmelnitsky}}\ and\ \bibinfo {author} {\bibfnamefont {V.}~\bibnamefont
  {Rubakov}},\ }\href {\doibase 10.1088/1475-7516/2014/02/019} {\bibfield
  {journal} {\bibinfo  {journal} {JCAP}\ }\textbf {\bibinfo {volume} {1402}},\
  \bibinfo {pages} {019} (\bibinfo {year} {2014})},\ \Eprint
  {http://arxiv.org/abs/1309.5888} {arXiv:1309.5888 [astro-ph.CO]} \BibitemShut
  {NoStop}%
\bibitem [{\citenamefont {{Porayko}}\ \emph {et~al.}(2018)\citenamefont
  {{Porayko}} \emph {et~al.}}]{poray18}%
  \BibitemOpen
  \bibfield  {author} {\bibinfo {author} {\bibfnamefont {N.~K.}\ \bibnamefont
  {{Porayko}}} \emph {et~al.},\ }\href {\doibase 10.1103/PhysRevD.98.102002}
  {\bibfield  {journal} {\bibinfo  {journal} {Phys. Rev. D}\ }\textbf {\bibinfo
  {volume} {98}},\ \bibinfo {eid} {102002} (\bibinfo {year} {2018})},\ \Eprint
  {http://arxiv.org/abs/1810.03227} {arXiv:1810.03227 [astro-ph.CO]}
  \BibitemShut {NoStop}%
\bibitem [{\citenamefont {Sokoloff}\ \emph {et~al.}(1998)\citenamefont
  {Sokoloff}, \citenamefont {Bykov}, \citenamefont {Shukurov}, \citenamefont
  {Berkhuijsen}, \citenamefont {Beck},\ and\ \citenamefont {Poezd}}]{sokol98}%
  \BibitemOpen
  \bibfield  {author} {\bibinfo {author} {\bibfnamefont {D.}~\bibnamefont
  {Sokoloff}}, \bibinfo {author} {\bibfnamefont {A.}~\bibnamefont {Bykov}},
  \bibinfo {author} {\bibfnamefont {A.}~\bibnamefont {Shukurov}}, \bibinfo
  {author} {\bibfnamefont {E.}~\bibnamefont {Berkhuijsen}}, \bibinfo {author}
  {\bibfnamefont {R.}~\bibnamefont {Beck}}, \ and\ \bibinfo {author}
  {\bibfnamefont {A.}~\bibnamefont {Poezd}},\ }\href {\doibase
  10.1046/j.1365-8711.1998.01782.x} {\bibfield  {journal} {\bibinfo  {journal}
  {Mon. Not. R. Astron. Soc.}\ }\textbf {\bibinfo {volume} {299}},\ \bibinfo
  {pages} {189} (\bibinfo {year} {1998})}\BibitemShut {NoStop}%
\bibitem [{\citenamefont {O'Sullivan}\ \emph {et~al.}(2012)\citenamefont
  {O'Sullivan}, \citenamefont {Brown}, \citenamefont {Robishaw}, \citenamefont
  {Schnitzeler}, \citenamefont {McClure-Griffiths}, \citenamefont {Feain},
  \citenamefont {Taylor}, \citenamefont {Gaensler}, \citenamefont {Landecker},
  \citenamefont {Harvey-Smith},\ and\ \citenamefont {Carretti}}]{sulli12}%
  \BibitemOpen
  \bibfield  {author} {\bibinfo {author} {\bibfnamefont {S.}~\bibnamefont
  {O'Sullivan}}, \bibinfo {author} {\bibfnamefont {S.}~\bibnamefont {Brown}},
  \bibinfo {author} {\bibfnamefont {T.}~\bibnamefont {Robishaw}}, \bibinfo
  {author} {\bibfnamefont {D.}~\bibnamefont {Schnitzeler}}, \bibinfo {author}
  {\bibfnamefont {N.}~\bibnamefont {McClure-Griffiths}}, \bibinfo {author}
  {\bibfnamefont {I.}~\bibnamefont {Feain}}, \bibinfo {author} {\bibfnamefont
  {A.}~\bibnamefont {Taylor}}, \bibinfo {author} {\bibfnamefont
  {B.}~\bibnamefont {Gaensler}}, \bibinfo {author} {\bibfnamefont
  {T.}~\bibnamefont {Landecker}}, \bibinfo {author} {\bibfnamefont
  {L.}~\bibnamefont {Harvey-Smith}}, \ and\ \bibinfo {author} {\bibfnamefont
  {E.}~\bibnamefont {Carretti}},\ }\href {\doibase
  10.1111/j.1365-2966.2012.20554.x} {\bibfield  {journal} {\bibinfo  {journal}
  {Mon. Not. R. Astron. Soc.}\ }\textbf {\bibinfo {volume} {421}},\ \bibinfo
  {pages} {3300} (\bibinfo {year} {2012})},\ \Eprint
  {http://arxiv.org/abs/1201.3161} {arXiv:1201.3161} \BibitemShut {NoStop}%
\bibitem [{\citenamefont {{O'Sullivan}}\ \emph {et~al.}(2017)\citenamefont
  {{O'Sullivan}}, \citenamefont {{Purcell}}, \citenamefont {{Anderson}},
  \citenamefont {{Farnes}}, \citenamefont {{Sun}},\ and\ \citenamefont
  {{Gaensler}}}]{sulli17}%
  \BibitemOpen
  \bibfield  {author} {\bibinfo {author} {\bibfnamefont {S.~P.}\ \bibnamefont
  {{O'Sullivan}}}, \bibinfo {author} {\bibfnamefont {C.~R.}\ \bibnamefont
  {{Purcell}}}, \bibinfo {author} {\bibfnamefont {C.~S.}\ \bibnamefont
  {{Anderson}}}, \bibinfo {author} {\bibfnamefont {J.~S.}\ \bibnamefont
  {{Farnes}}}, \bibinfo {author} {\bibfnamefont {X.~H.}\ \bibnamefont {{Sun}}},
  \ and\ \bibinfo {author} {\bibfnamefont {B.~M.}\ \bibnamefont {{Gaensler}}},\
  }\href {\doibase 10.1093/mnras/stx1133} {\bibfield  {journal} {\bibinfo
  {journal} {Mon. Not. R. Astron. Soc.}\ }\textbf {\bibinfo {volume} {469}},\
  \bibinfo {pages} {4034} (\bibinfo {year} {2017})},\ \Eprint
  {http://arxiv.org/abs/1705.00102} {arXiv:1705.00102 [astro-ph.GA]}
  \BibitemShut {NoStop}%
\bibitem [{\citenamefont {Agrawal}\ \emph {et~al.}(2019)\citenamefont
  {Agrawal}, \citenamefont {Hook},\ and\ \citenamefont {Huang}}]{agraw19}%
  \BibitemOpen
  \bibfield  {author} {\bibinfo {author} {\bibfnamefont {P.}~\bibnamefont
  {Agrawal}}, \bibinfo {author} {\bibfnamefont {A.}~\bibnamefont {Hook}}, \
  and\ \bibinfo {author} {\bibfnamefont {J.}~\bibnamefont {Huang}},\
  }\href@noop {} {\  (\bibinfo {year} {2019})},\ \Eprint
  {http://arxiv.org/abs/1912.02823} {arXiv:1912.02823 [astro-ph.CO]}
  \BibitemShut {NoStop}%
\bibitem [{\citenamefont {Liu}\ \emph {et~al.}(2020)\citenamefont {Liu},
  \citenamefont {Smoot},\ and\ \citenamefont {Zhao}}]{liu20}%
  \BibitemOpen
  \bibfield  {author} {\bibinfo {author} {\bibfnamefont {T.}~\bibnamefont
  {Liu}}, \bibinfo {author} {\bibfnamefont {G.}~\bibnamefont {Smoot}}, \ and\
  \bibinfo {author} {\bibfnamefont {Y.}~\bibnamefont {Zhao}},\ }\href {\doibase
  10.1103/PhysRevD.101.063012} {\bibfield  {journal} {\bibinfo  {journal}
  {Phys. Rev. D}\ }\textbf {\bibinfo {volume} {101}},\ \bibinfo {pages}
  {063012} (\bibinfo {year} {2020})},\ \Eprint
  {http://arxiv.org/abs/1901.10981} {arXiv:1901.10981 [astro-ph.CO]}
  \BibitemShut {NoStop}%
\bibitem [{\citenamefont {{Schwarz}}\ \emph {et~al.}(2020)\citenamefont
  {{Schwarz}}, \citenamefont {{Goswami}},\ and\ \citenamefont
  {{Basu}}}]{schwarz20}%
  \BibitemOpen
  \bibfield  {author} {\bibinfo {author} {\bibfnamefont {D.~J.}\ \bibnamefont
  {{Schwarz}}}, \bibinfo {author} {\bibfnamefont {J.}~\bibnamefont
  {{Goswami}}}, \ and\ \bibinfo {author} {\bibfnamefont {A.}~\bibnamefont
  {{Basu}}},\ }\href@noop {} {\bibfield  {journal} {\bibinfo  {journal} {arXiv
  e-prints}\ ,\ \bibinfo {eid} {arXiv:2003.10205}} (\bibinfo {year} {2020})},\
  \Eprint {http://arxiv.org/abs/2003.10205} {arXiv:2003.10205 [hep-ph]}
  \BibitemShut {NoStop}%
\bibitem [{\citenamefont {{Myers}}\ \emph {et~al.}(1999)\citenamefont
  {{Myers}}, \citenamefont {{Rusin}}, \citenamefont {{Fassnacht}},
  \citenamefont {{Bland ford}}, \citenamefont {{Pearson}}, \citenamefont
  {{Readhead}}, \citenamefont {{Jackson}}, \citenamefont {{Browne}},
  \citenamefont {{Marlow}}, \citenamefont {{Wilkinson}}, \citenamefont
  {{Koopmans}},\ and\ \citenamefont {{de Bruyn}}}]{myers99}%
  \BibitemOpen
  \bibfield  {author} {\bibinfo {author} {\bibfnamefont {S.~T.}\ \bibnamefont
  {{Myers}}}, \bibinfo {author} {\bibfnamefont {D.}~\bibnamefont {{Rusin}}},
  \bibinfo {author} {\bibfnamefont {C.~D.}\ \bibnamefont {{Fassnacht}}},
  \bibinfo {author} {\bibfnamefont {R.~D.}\ \bibnamefont {{Bland ford}}},
  \bibinfo {author} {\bibfnamefont {T.~J.}\ \bibnamefont {{Pearson}}}, \bibinfo
  {author} {\bibfnamefont {A.~C.~S.}\ \bibnamefont {{Readhead}}}, \bibinfo
  {author} {\bibfnamefont {N.}~\bibnamefont {{Jackson}}}, \bibinfo {author}
  {\bibfnamefont {I.~W.~A.}\ \bibnamefont {{Browne}}}, \bibinfo {author}
  {\bibfnamefont {D.~R.}\ \bibnamefont {{Marlow}}}, \bibinfo {author}
  {\bibfnamefont {P.~N.}\ \bibnamefont {{Wilkinson}}}, \bibinfo {author}
  {\bibfnamefont {L.~V.~E.}\ \bibnamefont {{Koopmans}}}, \ and\ \bibinfo
  {author} {\bibfnamefont {A.~G.}\ \bibnamefont {{de Bruyn}}},\ }\href
  {\doibase 10.1086/300875} {\bibfield  {journal} {\bibinfo  {journal} {Astron.
  J.}\ }\textbf {\bibinfo {volume} {117}},\ \bibinfo {pages} {2565} (\bibinfo
  {year} {1999})},\ \Eprint {http://arxiv.org/abs/astro-ph/9905043}
  {arXiv:astro-ph/9905043 [astro-ph]} \BibitemShut {NoStop}%
\bibitem [{\citenamefont {{Rusin}}\ \emph {et~al.}(2002)\citenamefont
  {{Rusin}}, \citenamefont {{Norbury}}, \citenamefont {{Biggs}}, \citenamefont
  {{Marlow}}, \citenamefont {{Jackson}}, \citenamefont {{Browne}},
  \citenamefont {{Wilkinson}},\ and\ \citenamefont {{Myers}}}]{rusin02}%
  \BibitemOpen
  \bibfield  {author} {\bibinfo {author} {\bibfnamefont {D.}~\bibnamefont
  {{Rusin}}}, \bibinfo {author} {\bibfnamefont {M.}~\bibnamefont {{Norbury}}},
  \bibinfo {author} {\bibfnamefont {A.~D.}\ \bibnamefont {{Biggs}}}, \bibinfo
  {author} {\bibfnamefont {D.~R.}\ \bibnamefont {{Marlow}}}, \bibinfo {author}
  {\bibfnamefont {N.~J.}\ \bibnamefont {{Jackson}}}, \bibinfo {author}
  {\bibfnamefont {I.~W.~A.}\ \bibnamefont {{Browne}}}, \bibinfo {author}
  {\bibfnamefont {P.~N.}\ \bibnamefont {{Wilkinson}}}, \ and\ \bibinfo {author}
  {\bibfnamefont {S.~T.}\ \bibnamefont {{Myers}}},\ }\href {\doibase
  10.1046/j.1365-8711.2002.05043.x} {\bibfield  {journal} {\bibinfo  {journal}
  {Mon. Not. R. Astron. Soc.}\ }\textbf {\bibinfo {volume} {330}},\ \bibinfo
  {pages} {205} (\bibinfo {year} {2002})},\ \Eprint
  {http://arxiv.org/abs/astro-ph/0110099} {astro-ph/0110099} \BibitemShut
  {NoStop}%
\bibitem [{Note1()}]{Note1}%
  \BibitemOpen
  \bibinfo {note} {Note that, \cite{rusin02} computed $\Delta t_{\protect \rm
  obs} =$ 39.7 day for the assumed Hubble-Lema\^itre constant $H_0 =
  100~\protect \rm km\protect \tmspace +\thinmuskip {.1667em}s^{-1}\protect
  \tmspace +\thinmuskip {.1667em}Mpc^{-1}$. Here, we have corrected the
  estimated time delay using $H_0 = 67.4~\protect \rm km\protect \tmspace
  +\thinmuskip {.1667em}s^{-1}\protect \tmspace +\thinmuskip {.1667em}Mpc^{-1}$
  obtained from the {\protect \it Planck} satellite's data \cite{planck18VI}
  and used $\Delta t_{\protect \rm obs} =$ 27.8~day in our
  calculations.}\BibitemShut {Stop}%
\bibitem [{\citenamefont {{Mao}}\ \emph {et~al.}(2017)\citenamefont {{Mao}},
  \citenamefont {{Carilli}}, \citenamefont {{Gaensler}}, \citenamefont
  {{Wucknitz}}, \citenamefont {{Keeton}}, \citenamefont {{Basu}}, \citenamefont
  {{Beck}}, \citenamefont {{Kronberg}},\ and\ \citenamefont
  {{Zweibel}}}]{mao17}%
  \BibitemOpen
  \bibfield  {author} {\bibinfo {author} {\bibfnamefont {S.~A.}\ \bibnamefont
  {{Mao}}}, \bibinfo {author} {\bibfnamefont {C.}~\bibnamefont {{Carilli}}},
  \bibinfo {author} {\bibfnamefont {B.~M.}\ \bibnamefont {{Gaensler}}},
  \bibinfo {author} {\bibfnamefont {O.}~\bibnamefont {{Wucknitz}}}, \bibinfo
  {author} {\bibfnamefont {C.}~\bibnamefont {{Keeton}}}, \bibinfo {author}
  {\bibfnamefont {A.}~\bibnamefont {{Basu}}}, \bibinfo {author} {\bibfnamefont
  {R.}~\bibnamefont {{Beck}}}, \bibinfo {author} {\bibfnamefont {P.~P.}\
  \bibnamefont {{Kronberg}}}, \ and\ \bibinfo {author} {\bibfnamefont
  {E.}~\bibnamefont {{Zweibel}}},\ }\href {\doibase
  https://doi.org/10.1038/s41550-017-0218-x} {\bibfield  {journal} {\bibinfo
  {journal} {Nature Astronomy}\ }\textbf {\bibinfo {volume} {1}},\ \bibinfo
  {pages} {621} (\bibinfo {year} {2017})}\BibitemShut {NoStop}%
\bibitem [{fis()}]{fisher93}%
  \BibitemOpen
  \href@noop {} {}\bibinfo {note} {N. Fisher, Statistical Analysis of Circular
  Data, Cambridge: Cambridge University Press (1993).}\BibitemShut {Stop}%
\bibitem [{mar()}]{mardia00}%
  \BibitemOpen
  \href@noop {} {}\bibinfo {note} {K. V. Mardia, P. E. Jupp, (2000),
  Directional Statistics, John Wiley \& Sons, Inc., New York
  (2000).}\BibitemShut {Stop}%
\bibitem [{\citenamefont {{Lyutikov}}(2017)}]{lyuti17}%
  \BibitemOpen
  \bibfield  {author} {\bibinfo {author} {\bibfnamefont {M.}~\bibnamefont
  {{Lyutikov}}},\ }\href {\doibase 10.1103/PhysRevD.95.124003} {\bibfield
  {journal} {\bibinfo  {journal} {Phys. Rev. D}\ }\textbf {\bibinfo {volume}
  {95}},\ \bibinfo {eid} {124003} (\bibinfo {year} {2017})},\ \Eprint
  {http://arxiv.org/abs/1703.00601} {arXiv:1703.00601 [gr-qc]} \BibitemShut
  {NoStop}%
\bibitem [{\citenamefont {{Rumbaugh}}\ \emph {et~al.}(2015)\citenamefont
  {{Rumbaugh}}, \citenamefont {{Fassnacht}}, \citenamefont {{McKean}},
  \citenamefont {{Koopmans}}, \citenamefont {{Auger}},\ and\ \citenamefont
  {{Suyu}}}]{rumba15}%
  \BibitemOpen
  \bibfield  {author} {\bibinfo {author} {\bibfnamefont {N.}~\bibnamefont
  {{Rumbaugh}}}, \bibinfo {author} {\bibfnamefont {C.~D.}\ \bibnamefont
  {{Fassnacht}}}, \bibinfo {author} {\bibfnamefont {J.~P.}\ \bibnamefont
  {{McKean}}}, \bibinfo {author} {\bibfnamefont {L.~V.~E.}\ \bibnamefont
  {{Koopmans}}}, \bibinfo {author} {\bibfnamefont {M.~W.}\ \bibnamefont
  {{Auger}}}, \ and\ \bibinfo {author} {\bibfnamefont {S.~H.}\ \bibnamefont
  {{Suyu}}},\ }\href {\doibase 10.1093/mnras/stv672} {\bibfield  {journal}
  {\bibinfo  {journal} {Mon. Not. R. Astron. Soc.}\ }\textbf {\bibinfo {volume}
  {450}},\ \bibinfo {pages} {1042} (\bibinfo {year} {2015})},\ \Eprint
  {http://arxiv.org/abs/1410.6557} {arXiv:1410.6557 [astro-ph.CO]} \BibitemShut
  {NoStop}%
\bibitem [{Note2()}]{Note2}%
  \BibitemOpen
  \bibinfo {note} {Here we have made the approximation $\langle \rho_{\rm a, em}^{1/2} \rangle \approx \langle \rho_{\rm a, em} \rangle^{1/2}$. This is valid when $\langle \rho_{\rm a, em}\rangle$ is larger than the dispersion of $\rho_
{\rm a, em}$}\BibitemShut {Stop}%
\bibitem [{\citenamefont {{Koopmans}}\ \emph {et~al.}(2006)\citenamefont
  {{Koopmans}}, \citenamefont {{Treu}}, \citenamefont {{Bolton}}, \citenamefont
  {{Burles}},\ and\ \citenamefont {{Moustakas}}}]{koopm06}%
  \BibitemOpen
  \bibfield  {author} {\bibinfo {author} {\bibfnamefont {L.~V.~E.}\
  \bibnamefont {{Koopmans}}}, \bibinfo {author} {\bibfnamefont
  {T.}~\bibnamefont {{Treu}}}, \bibinfo {author} {\bibfnamefont {A.~S.}\
  \bibnamefont {{Bolton}}}, \bibinfo {author} {\bibfnamefont {S.}~\bibnamefont
  {{Burles}}}, \ and\ \bibinfo {author} {\bibfnamefont {L.~A.}\ \bibnamefont
  {{Moustakas}}},\ }\href {\doibase 10.1086/505696} {\bibfield  {journal}
  {\bibinfo  {journal} {Astroph. J.}\ }\textbf {\bibinfo {volume} {649}},\
  \bibinfo {pages} {599} (\bibinfo {year} {2006})},\ \Eprint
  {http://arxiv.org/abs/astro-ph/0601628} {arXiv:astro-ph/0601628 [astro-ph]}
  \BibitemShut {NoStop}%
\bibitem [{\citenamefont {{Lyskova}}\ \emph {et~al.}(2018)\citenamefont
  {{Lyskova}}, \citenamefont {{Churazov}},\ and\ \citenamefont
  {{Naab}}}]{lysko18}%
  \BibitemOpen
  \bibfield  {author} {\bibinfo {author} {\bibfnamefont {N.}~\bibnamefont
  {{Lyskova}}}, \bibinfo {author} {\bibfnamefont {E.}~\bibnamefont
  {{Churazov}}}, \ and\ \bibinfo {author} {\bibfnamefont {T.}~\bibnamefont
  {{Naab}}},\ }\href {\doibase 10.1093/mnras/sty018} {\bibfield  {journal}
  {\bibinfo  {journal} {Mon. Not. R. Astron. Soc.}\ }\textbf {\bibinfo {volume}
  {475}},\ \bibinfo {pages} {2403} (\bibinfo {year} {2018})},\ \Eprint
  {http://arxiv.org/abs/1711.01123} {arXiv:1711.01123 [astro-ph.GA]}
  \BibitemShut {NoStop}%
\bibitem [{\citenamefont {{Lyskova}}\ \emph {et~al.}(2014)\citenamefont
  {{Lyskova}}, \citenamefont {{Churazov}}, \citenamefont {{Moiseev}},
  \citenamefont {{Sil'chenko}},\ and\ \citenamefont {{Zhuravleva}}}]{lysko14}%
  \BibitemOpen
  \bibfield  {author} {\bibinfo {author} {\bibfnamefont {N.}~\bibnamefont
  {{Lyskova}}}, \bibinfo {author} {\bibfnamefont {E.}~\bibnamefont
  {{Churazov}}}, \bibinfo {author} {\bibfnamefont {A.}~\bibnamefont
  {{Moiseev}}}, \bibinfo {author} {\bibfnamefont {O.}~\bibnamefont
  {{Sil'chenko}}}, \ and\ \bibinfo {author} {\bibfnamefont {I.}~\bibnamefont
  {{Zhuravleva}}},\ }\href {\doibase 10.1093/mnras/stu717} {\bibfield
  {journal} {\bibinfo  {journal} {Mon. Not. R. Astron. Soc.}\ }\textbf
  {\bibinfo {volume} {441}},\ \bibinfo {pages} {2013} (\bibinfo {year}
  {2014})},\ \Eprint {http://arxiv.org/abs/1404.2729} {arXiv:1404.2729
  [astro-ph.GA]} \BibitemShut {NoStop}%
\bibitem [{\citenamefont {{Lovell}}\ \emph {et~al.}(2018)\citenamefont
  {{Lovell}}, \citenamefont {{Pillepich}}, \citenamefont {{Genel}},
  \citenamefont {{Nelson}}, \citenamefont {{Springel}}, \citenamefont
  {{Pakmor}}, \citenamefont {{Marinacci}}, \citenamefont {{Weinberger}},
  \citenamefont {{Torrey}}, \citenamefont {{Vogelsberger}}, \citenamefont
  {{Alabi}},\ and\ \citenamefont {{Hernquist}}}]{lovel18}%
  \BibitemOpen
  \bibfield  {author} {\bibinfo {author} {\bibfnamefont {M.~R.}\ \bibnamefont
  {{Lovell}}}, \bibinfo {author} {\bibfnamefont {A.}~\bibnamefont
  {{Pillepich}}}, \bibinfo {author} {\bibfnamefont {S.}~\bibnamefont
  {{Genel}}}, \bibinfo {author} {\bibfnamefont {D.}~\bibnamefont {{Nelson}}},
  \bibinfo {author} {\bibfnamefont {V.}~\bibnamefont {{Springel}}}, \bibinfo
  {author} {\bibfnamefont {R.}~\bibnamefont {{Pakmor}}}, \bibinfo {author}
  {\bibfnamefont {F.}~\bibnamefont {{Marinacci}}}, \bibinfo {author}
  {\bibfnamefont {R.}~\bibnamefont {{Weinberger}}}, \bibinfo {author}
  {\bibfnamefont {P.}~\bibnamefont {{Torrey}}}, \bibinfo {author}
  {\bibfnamefont {M.}~\bibnamefont {{Vogelsberger}}}, \bibinfo {author}
  {\bibfnamefont {A.}~\bibnamefont {{Alabi}}}, \ and\ \bibinfo {author}
  {\bibfnamefont {L.}~\bibnamefont {{Hernquist}}},\ }\href {\doibase
  10.1093/mnras/sty2339} {\bibfield  {journal} {\bibinfo  {journal} {Mon. Not.
  R. Astron. Soc.}\ }\textbf {\bibinfo {volume} {481}},\ \bibinfo {pages}
  {1950} (\bibinfo {year} {2018})},\ \Eprint {http://arxiv.org/abs/1801.10170}
  {arXiv:1801.10170 [astro-ph.GA]} \BibitemShut {NoStop}%
\bibitem [{\citenamefont {{McKean}}\ \emph {et~al.}(2015)\citenamefont
  {{McKean}}, \citenamefont {{Jackson}}, \citenamefont {{Vegetti}},
  \citenamefont {{Rybak}}, \citenamefont {{Serjeant}}, \citenamefont
  {{Koopmans}}, \citenamefont {{Metcalf}}, \citenamefont {{Fassnacht}},
  \citenamefont {{Marshall}},\ and\ \citenamefont
  {{Pandey-Pommier}}}]{mckea15}%
  \BibitemOpen
  \bibfield  {author} {\bibinfo {author} {\bibfnamefont {J.}~\bibnamefont
  {{McKean}}}, \bibinfo {author} {\bibfnamefont {N.}~\bibnamefont {{Jackson}}},
  \bibinfo {author} {\bibfnamefont {S.}~\bibnamefont {{Vegetti}}}, \bibinfo
  {author} {\bibfnamefont {M.}~\bibnamefont {{Rybak}}}, \bibinfo {author}
  {\bibfnamefont {S.}~\bibnamefont {{Serjeant}}}, \bibinfo {author}
  {\bibfnamefont {L.~V.~E.}\ \bibnamefont {{Koopmans}}}, \bibinfo {author}
  {\bibfnamefont {R.~B.}\ \bibnamefont {{Metcalf}}}, \bibinfo {author}
  {\bibfnamefont {C.}~\bibnamefont {{Fassnacht}}}, \bibinfo {author}
  {\bibfnamefont {P.~J.}\ \bibnamefont {{Marshall}}}, \ and\ \bibinfo {author}
  {\bibfnamefont {M.}~\bibnamefont {{Pandey-Pommier}}},\ }in\ \href@noop {}
  {\emph {\bibinfo {booktitle} {Advancing Astrophysics with the Square
  Kilometre Array (AASKA14)}}}\ (\bibinfo {year} {2015})\ p.~\bibinfo {pages}
  {84},\ \Eprint {http://arxiv.org/abs/1502.03362} {arXiv:1502.03362
  [astro-ph.GA]} \BibitemShut {NoStop}%
\bibitem [{bra()}]{braun}%
  \BibitemOpen
  \href@noop {} {}\bibinfo {note} {R. Braun, et al., SKA document number
  SKA-TEL-SKO-0000818, (2017),
  \url{https://astronomers.skatelescope.org/wp-content/uploads/2017/10/SKA-TEL-SKO-0000818-01_SKA1_Science_Perform.pdf}}\BibitemShut
  {NoStop}%
\bibitem [{\citenamefont {{Planck Collaboration VI}}(2018)}]{planck18VI}%
  \BibitemOpen
  \bibfield  {author} {\bibinfo {author} {\bibnamefont {{Planck Collaboration
  VI}}},\ }\href@noop {} {\bibfield  {journal} {\bibinfo  {journal} {}
  } (\bibinfo {year} {2018})} \Eprint
  {http://arxiv.org/abs/1807.06209} {arXiv:1807.06209}
  \BibitemShut {NoStop}%
\end{thebibliography}

%

\end{document}